*Perspective*

# Cybersecurity in Power Grids: Challenges and Opportunities

Tim Krause [1], Raphael Ernst [1], Benedikt Klaer [2,3], Immanuel Hacker [2,3] and Martin Henze [1,*]

1. Cyber Analysis & Defense, Fraunhofer FKIE, 53343 Wachtberg, Germany; tim.krause@fkie.fraunhofer.de (T.K.); raphael.ernst@fkie.fraunhofer.de (R.E.)
2. Digital Energy, Fraunhofer FIT, 52056 Aachen, Germany; benedikt.klaer@fit.fraunhofer.de (B.K.); immanuel.hacker@fit.fraunhofer.de (I.H.)
3. High Voltage Equipment and Grids, Digitalisation and Power Economics, RWTH Aachen University, 52056 Aachen, Germany
* Correspondence: martin.henze@fkie.fraunhofer.de

**Abstract:** Increasing volatilities within power transmission and distribution force power grid operators to amplify their use of communication infrastructure to monitor and control their grid. The resulting increase in communication creates a larger attack surface for malicious actors. Indeed, cyber attacks on power grids have already succeeded in causing temporary, large-scale blackouts in the recent past. In this paper, we analyze the communication infrastructure of power grids to derive resulting fundamental challenges of power grids with respect to cybersecurity. Based on these challenges, we identify a broad set of resulting attack vectors and attack scenarios that threaten the security of power grids. To address these challenges, we propose to rely on a defense-in-depth strategy, which encompasses measures for (i) device and application security, (ii) network security, and (iii) physical security, as well as (iv) policies, procedures, and awareness. For each of these categories, we distill and discuss a comprehensive set of state-of-the art approaches, as well as identify further opportunities to strengthen cybersecurity in interconnected power grids.

## 1. Introduction

Historically, power grids have grown from simple, localized grids to large, physically wide-spread grids, often spanning multiple nations or even whole continents [1]. Despite its importance to modern society, the energy sector has adapted slower than other industries to digital technology due to its size and need for high system availability. This is also reflected by a legislative point of view, with power grids being classified as critical infrastructure in many countries [2–5]. Because of the need for more efficiency, digital technology gets more widespread, and new technologies in the energy grid heavily rely on high-frequency monitoring cycles and adaptation to bottlenecks in the grid [6]. This trend is boosted with the rise of renewable energy (ranging from large off-shore wind farms matching the power generation of traditional power plants to a single household feeding solar energy into the grid) [7,8], which leads to power generation becoming more distributed and, thus, less reliable, resulting in difficult to organize transmission and distribution of energy [9,10].

This results in a less controllable situation than in the past, when only a small number of bulk power production plants were required. While increasing power demands can be satisfied with more traditional or renewable power plants, the grid itself needs to support the transportation of the generated power. However, extending the grid by adding new lines is prohibitively expensive and often slowed down, e.g., by regulations or resistance of residents. Fortunately, digital technology can aid in better utilizing the existing grid, leading to an increased deployment of digital technology to control, monitor, and maintain transmission and distribution of power [2,11].

This increasing use of digital technology requires more and more networking capabilities and connects previously isolated components with larger communication networks [12–14], resulting in a large variety of new dataflows [15]. The resulting increasing interconnection of power grids, which constitute critical infrastructure [2–5,16] requiring special protection, raises severe security concerns [17]: Protocols and systems originally developed for power grids were not designed with security in mind. Yet, these systems are still used alongside modern technology and increasingly exposed to outside networks, such as the Internet. Likewise, the increasing use of digital and decentralized technology provides a larger attack surface [18–20]. Indeed, different cyber attacks have successfully targeted essential parts of the power grid [21,22]. Resulting disruptions and wide-scale outages of electrical power have extensive social and economic consequences [23].

*Contributions*

This paper specifically targets the security challenges originating from the increasing interconnection of power grids, especially at the transmission and distribution level. To this end, we motivate the need for cybersecurity when operating power grids and outline promising approaches to provide security at different levels of abstraction. More specifically, our contributions in this paper are:

1. We provide a high-level overview over the communication infrastructure of power grids and derive resulting fundamental challenges w.r.t. cybersecurity risks (Section 2).
2. As a foundation to secure power grids, we identify a comprehensive set of attack vectors and scenarios based on these security challenges (Section 3).
3. We distill and discuss promising approaches to provide security for interconnected power grids to protect against serious attack vectors and scenarios (Section 4).

Extending upon previous works on surveying various aspects of cybersecurity challenges in power grids and smart grid environments [16,24–26], this paper provides a unique perspective on cybersecurity challenges and opportunities in power grids by combining the viewpoints of cybersecurity and communication network researchers with those of electrical engineers and power grid operators. As such, our contributions are relevant both for cybersecurity researchers, who are usually not familiar with the processes and terminology of electrical engineering, as well as electrical engineers and power grid operators, who have a decent understanding of power grids but are often unaware of the extent and specifics of cybersecurity challenges.

## 2. Communication Infrastructure of Power Grids and Resulting Security Challenges

With the increasing digitalization of power grids, operators are confronted with rapid changes in the amount of communication necessary and the means through which this communication is conducted. As a result, more and more communication is introduced to power grids. In the following, we first describe the communication infrastructure of power grids, before focusing on the fundamental security challenges resulting from an increasing interconnection of power grids.

We use the term *grid* exclusively to refer to the power grid and the term *network* for digital communication networks. Furthermore, the information contained in this paper is mainly focused on European power grids. However, many of the proposed changes to cybersecurity also apply to other regions. In this context, we define the transmission of power as the transportation of energy over long distances (e.g., between distant cities, not within a city) and the distribution of power as the transportation on a local scale, such as a single city or small region. The transmission of power is overseen by *transmission system operators* (TSO), while the distribution of power is carried out by *distribution system operators* (DSO).

*2.1. Communication Infrastructure of Power Grids*

To illustrate the typical communication infrastructure found in power grids, we provide a simplified view of a grid operator network, i.e., a TSO or DSO, in Figure 1. Typical to the network of a grid operator is the separation between *office network* and *process control network* (PCN). The office network is similar to any usual corporate network with, e.g., email traffic or data processing. Contrary, the PCN connects the control room of grid operation companies with their substations and field devices, typically using DNP3 [27] (North America and parts of Asia) or IEC 60870-5-104 [28] (rest of the world) as protocol. Resulting control messages are usually interpreted by a *programmable logic controller* (PLC) and then passed to the process layer. The control room typically contains a *human-machine interface* (HMI), a database (DB) server managing grid information, and a simulation server for pre-computing the effects of grid changes. Furthermore, the control room is connected to multiple substations (each containing at least a gateway, an HMI, and multiple PLCs) and can be coupled with other TSO/DSO SCADA systems for mutual control.

Data exchange between office network and PCN should only be handled through a dedicated data exchange server, where every file is checked for malware before being passed through. Sometimes, though, as seen in the Ukraine attacks [21], there are other communication channels, such as VPNs, which allow direct communication between the office network and the PCN or remote maintenance lines for vendors or contractors.

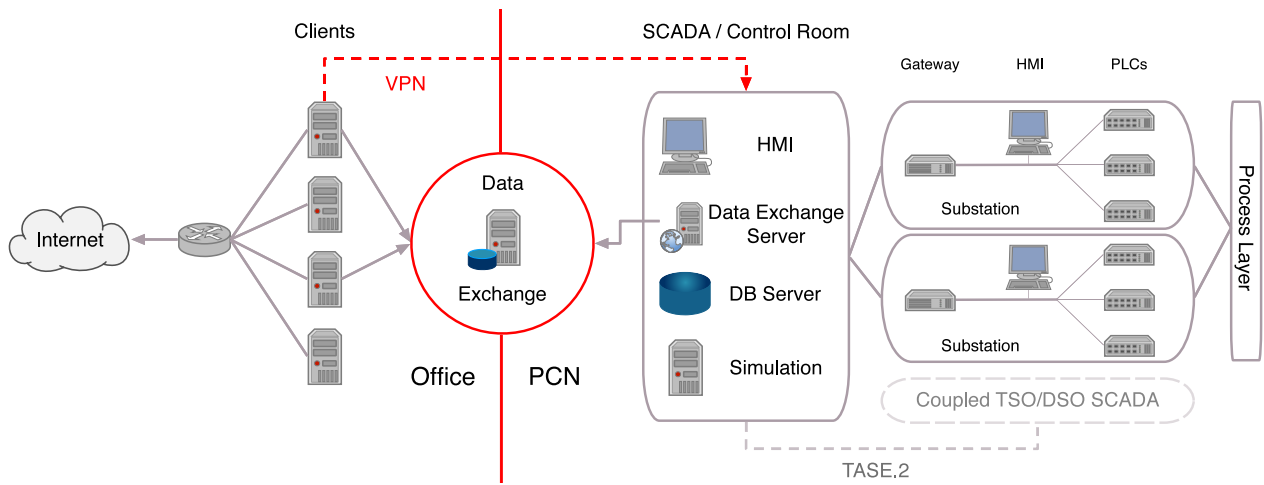

**Figure 1.** A simplified view of a TSO/DSO network, separated into office network (connected to the Internet, typical data processing tasks) and process control network (SCADA traffic, connecting the control room with substations and field devices). While data should only be exchanged through a dedicated data exchange server, in reality, these networks are often interconnected, e.g., with VPNs.

*2.2. Fundamental Cybersecurity Challenges*

While the increasing digitalization of power grids is necessary to deal with changing power demands and generation, it raises fundamental security challenges, especially when considering that power grids are an essential part of critical infrastructure [2–5]. In the following, we introduce and highlight the most important challenges or mechanisms in electrical power grids impacting cybersecurity.

2.2.1. CIA Triad: Availability Is Key

The triad of confidentiality, integrity, and availability (CIA) [29] as the fundamental concept of information security has to be interpreted slightly differently in the energy sector. In traditional cybersecurity, it is generally preferable to ensure confidentiality and integrity and possibly sacrifice (some) availability. In power grids, however, availability is by far the most important measure of the triad, as the consequences of downtime can be severe [23]. The longer a blackout lasts and the more of the grid is affected, the harder it is to rebuild the grid [30]. To illustrate the race for availability, the German power grid had an end-consumer availability of 99.9995% in 2017 [31], compared to the allegedly highly available Google services with 99.978% (no scheduled downtime) [32]. When measured in time, the power grid has a more than 45-fold higher availability. As availability is the most important measure in power grids, any measures ensuring the confidentiality and integrity of systems should never interfere with the availability of power delivery.

2.2.2. Balancing Generation and Consumption

Electrical power grids rely on a stable grid frequency of either 50 Hz or 60 Hz due to the use of alternating current. The frequency is only stable if power generation and consumption are at an equilibrium. If more power is generated than consumed, the frequency rises, and vice versa.

The operation reserve ensures the equilibrium between production and consumption at any given time. In Europe, the operation reserve is separated in three different stages: primary operation reserve, secondary-reserve, and minute reserve. The primary control is a continuous frequency load control, where the controller is implemented as an droop control distributed to different power plants. Thereby, the amount of primary operation reserve is 3 GW which is the power rating of two mayor generation units (e.g., nuclear power plants). The secondary-reserve and minute-reserve will be activated if the frequency derivation holds on. More specifically, the secondary reserve has to be activated after 5 min of frequency derivation and will be replaced by the minute reserve after 15 min. These are both activated centrally by the grid operator in whose grid area the power deviation occurred. However, at deviations of 1 Hz, either some consumers or, possibly, multiple cities have to be disconnected from the grid (in case of decreased grid frequency) or power plants have to be downregulated to save the generators from damage [33,34].

The sensitive equilibrium between generation and consumption can be exploited by attackers, as they only need to control a comparably small amount of consumption or generation to use cascading effects within the grid to create a system-wide blackout [35,36], as power generation and consumption in the operating reserve have time delays.

2.2.3. Decentralization of Power Generation

The rise of renewable energy has empowered many individuals and companies to enter the energy sector [37]. For example, individual households can now feed their excess solar energy into the grid. Naturally, the security of their systems is not as tightly controlled as those of traditional energy companies. As a result, the hardware and software use by individuals to operate power generation are often not as secure as it should be or misconfigured [38], potentially impacting transmission and distribution in the grid.

Assuming a vulnerability in a large number of, e.g., solar installations is found, attackers may control the power fed into the grid. Consequently, an attacker can do considerable damage even when controlling only a comparably small amount of energy by exploiting cascading effects (see above) [37]. By controlling the feeding of power into the grid, an attacker could exploit these cascading effects with the goal to cause a system-wide blackout, impacting distribution and transmission [35,36].

2.2.4. No Security in Process Control Networks

Most devices used in power grids, such as protection devices or PLCs, are designed for multiple-decade use. Often, they are neither patched nor replaced. The most widespread protocols are DNP3 [27] (mostly used in North America and parts of Asia) and IEC 60870-5-104 [28] (predominantly used by the rest of the world). These protocols have been developed more than 20 years ago with no security concerns in mind [39]. Despite their wide-spread use by DSOs and TSOs, neither protocol supports basic security mechanisms, such as authentication or integrity protection, which have been taken for granted for years in other commercial sectors [40]. For example, IEC 60870-5-104 [28], as used by DSOs and TSOs in Europe and other parts of the world, is susceptible to man-in-the-middle, denial-of-service, replay, and spoofing attacks [41–43]. Similar observations have been made for the DNP3 [27] protocol used in North America and parts of Asia [44–46]. While the standard IEC 62351 provides additional cybersecurity concepts [47], it is usually not implemented.

As grid operators often use their own separated networks over dedicated physical cables, they could neglect further security mechanisms in the past, leading to networks which do not necessarily use cryptography, authentication, or integrity checks. Past attacks have shown that office networks (connected to the Internet) are often not sufficiently separated from the PCN, allowing attackers lateral movement between the two [21]. Once an attacker gains access to an unsecured PCN, simple tools enabling communication in the specific protocol may be used to control devices crucial for grid operation.

### 2.2.5. Difficulty of Physical Network Changes

Field devices in power grids have planned lifetimes which are measured not in years but decades [48]. Information technology has much faster development cycles, allowing the development of attacks for devices which may have to be in use for many more years. It is very unlikely that all energy companies will be able to always follow current security best-practices by simply exchanging devices for newer models, which, e.g., support more modern protocols, such as IEC 61850 [49].

If possible at all, modern security mechanisms have to be implemented in software only and run on the available hardware or be able to interface directly with the specialized devices used by grid operation companies, without affecting the availability of electricity supply. However, many devices in use may not have the computational power to support additional security functionality [50–54]. Even if certain devices are eventually exchanged or upgraded with new software, they are often required to support legacy protocols to be able to communicate with older devices still relying on these protocols. Consequently, distribution and transmission system operators suffer from the well-known problem of insecurity by inheritance [55].

### 2.2.6. Weakest Link Problem

For attacks to have devastating consequences, an attacker does not have to target the largest grid operator. As long as the victim of an attack has control over enough power to affect the grid frequency, the attacker can leverage cascading effects to affect the whole power grid. As smaller operators often lack the means to harden their systems as much as larger operators, such targets may be more attractive to attackers. Furthermore, attacks do not have to be limited to grid operators: An attacker controlling a larger number of consumer electronics, e.g., solar power cells, might still be able to influence the frequency within the grid [36].

As a result, there is a need to develop solutions which can be used by all relevant actors in interconnected power grids and are not only deployable by larger grid operators.

## 3. Attack Vectors and Scenarios

Practical cybersecurity in interconnected power grids is impacted by a diverse set of fundamental security challenges. As a foundation to overcome these challenges and, thus, provide security for distribution and transmission systems, we now identify attack vectors and attack scenarios that result from the fundamental security challenges. In the following, we first discuss the most important attack vectors before we present the attack scenarios enabled by these vectors.

### 3.1. Attack Vectors in Distribution and Transmission Grids

Attackers can leverage different attack vectors to compromise the network of a transmission or distribution system operator with the goal of causing a blackout or at least considerable disturbance in the power grid. To achieve this goal, an attacker will likely aim to compromise the PCN of the target system. From there, the attacker can compromise substations or field devices, potentially leading to a blackout. In the following, we discuss the most important attack vectors an attacker can exploit to access a PCN. We provide a summary of our classification of attack vectors in Table 1.

Table 1. Classification of attack vectors specific for the energy sector.

|  | Scope | Difficulty | Impact | Examples |
|---|---|---|---|---|
| **Lateral Movement** | Single Operator | High | High | [21,56,57] |
| **Physical Access** | Local | Medium | Medium | [24,58–60] |
| **Remote Maintenance Access** | Multiple Operators | High | High | [61–63] |
| **Third-Party Exploit** | Multiple Operators | High | Medium | [22,64,65] |
| **Overcoming Air Gap** | Local | High | Medium | [56,57] |
| **Insider Attack** | Single Operator | Low | High | [24,66] |
| **Cascading Effects** | Multiple Operators | High | High | [35,36] |

### 3.1.1. Lateral Movement from the Office Network

In the attacks on Ukrainian grid operators in 2015 [21], attackers gained access to the PCN through lateral movement from the office network (cf. Figure 1). Allowing communications between PCN-connected devices and the office network might be necessary, e.g., to transfer certain information, such as environment data between office network and control room. An attacker can comprise an office network, e.g., by sending spear phishing emails to certain employees or by exploiting vulnerabilities in applications, such as web browsers or office suites [67]. Once access to a machine in the office network has been gained, the attacker can passively listen for user credentials and search for, e.g., a VPN tunnel to the PCN. Lateral movement from the office network is one of the most dangerous attack vectors for power system operators. An attack would, however, be limited to a single network.

### 3.1.2. Physical Access

Energy providers often use their own dedicated cable networks for PCN communications (cf. Section 2.2.4). While this certainly provides an extra level of security, it does not offer any protection once an attacker has gained physical access to one device in the network. For example, substations are usually connected to the PCN and are controlled remotely without personnel being present. Therefore, an attacker can gain physical access to the network by breaking into a substation and then manually use available systems or connect its own device to the PCN [24]. Break-ins in substations already happen today [58,59], although mostly with the intention to steal and then sell copper cables [24,60]. Hence, such a theft could be used as camouflage by cyber attackers to deter grid operators from even looking for traces of a cyber attack. Such an attack could have a large influence on grid operations but does not scale well, as it necessitates physical presence at (possibly multiple) substations.

### 3.1.3. Remote Maintenance Access

Manufacturers of control room software and hardware usually have a maintenance contract with the grid operators using their systems. To be able to debug these systems remotely or to deploy software updates, these systems are typically equipped with some form of remote maintenance access [61]. Depending on the technology used and security measures in place, attackers may try to exploit this maintenance access to gain access to the PCN. Such maintenance access is typically hardened against cyber attacks. However, if a vulnerability is found, an attack could have a considerable impact at multiple operators and wide-ranging control over the PCN is likely. As an example, in 2014 ICS-CERT reported on a security breach at a public utility where the attacker used standard brute forcing techniques to gain access to a password-protected remote access [62]. The risk of this attack vector is further illustrated by an incident from 2013, in which attackers compromised a vendor of a large power producer in the U.S. and Canada to exfiltrate passwords potentially used for remote maintenance access [63].

3.1.4. Third-Party Exploit

Attack vectors are not limited to the premises of grid operators. In fact, exploiting third-parties, such as suppliers or subcontractors, is one of the most dangerous and hard to control attack vectors. For example, the actors behind the Dragonfly malware, which specifically targets energy supplies, compromised three different manufacturers of ICS equipment and inserted their malware into software available on the manufacturers' websites [64]. Likewise, for the attack on the Ukrainian power grid in 2016 [22], attackers compromised manufacturers of field devices and manipulated firmware update installers on the publicly available websites of the manufacturers. Employees at the affected DSO then downloaded and deployed these updates, consequently unknowingly installing the attackers' malware which allowed the attackers to gain access to PCN-connected devices. Moreover, in case the manufacturer of the control room software is compromised, attackers may have direct access to the remote maintenance access of multiple grid operators. This becomes especially relevant when relying on cloud resources [48,68,69]. In a different case, the turbine control system of a power company in the United State was unintentionally infected with a virus through a infected USB drive plugged in by a third-party technician during maintenance [65]. A special case of a third-party exploit would be a compromise of the supply chain during the manufacturing of field devices. If attackers are able to tamper with devices before they are installed at the grid operator, they might, e.g., install a covert channel for remote access. The impact of such an attack would be high and could result in extensive access to the PCN. A single exploit could be used to gain control over multiple grid operators.

3.1.5. Overcoming Air Gap

Even if PCNs are air-gapped, i.e., physically isolated from other networks, such as the office network to prevent lateral movement, attackers can still try to attack a PCN by strategically placing USB drives containing malware around a facility they are targeting. Consequently, curious or helpful employees may unknowingly compromise air-gapped systems by connecting such drives to devices in company networks. Such an attack would likely be local in scope with a medium impact, as attackers could only execute previously determined attacks. Because of the air gap, there will be no immediate feedback on the attacks' success to an adversary. Indeed, malware-infected USB drive are one suspected attack vector of Stuxnet [56,57]. Notably, air-gapped systems in areas with even stronger security requirements, such as military bases [70] or the international space station [71], have shown to be susceptible to such attacks, especially using USB drives.

3.1.6. Insider Attack

If an attacker already works within the energy sector or compromises an employee of a grid operator, the attacker might have direct access to the control room or field devices and could, therefore, directly control devices or introduce malware, even to air-gapped systems. Insider attacks are hard to predict and protect against. In the case of an insider, no further compromise of systems may be necessary, as the attacker has legitimate access to the PCN. Different examples of disgruntled employees misusing their authority have been reported by Brdiczka [66]. For example, a disgruntled employee caused a power outage during the 2002 Winter Olympics by knocking out a substation in Salt Lake City [24]. An insider attack can have a high impact on the grid, especially since insiders typically have decent knowledge of the inner workings of grids and potential security measures in place, thus being able to carefully pick their target.

3.1.7. Cascading Effects

Instead of having to compromise the network of one or more grid operators, an attacker may leverage cascading effects in the power grid to cause a power outage (cf. Section 2). For example, by remotely gaining control over a large number of consumer electronics, such as solar power cells [72], an attacker can take advantage of the mechanics of the oper-ating reserve to influence the frequency within the grid [36]. There are also companies that centrally manage many distributed solar or wind plants, e.g., within the scope of a virtual power plant [73]. A compromise at one of these service providers may allow attackers to leverage similar effects. An attack exploiting cascading effects could be global in scale with a medium to high impact (depending on the amount of power under the attackers' control) but has a high technical difficulty, as a comparably large number of devices has to be exploited and controlled simultaneously. However, already today, larger botnets, e.g., Conficker, Hajime, or WannaCry, control hundreds of thousands to millions of devices [74–77] making such attacks less unlikely than it appears. The problem might further exaggerate with the rise of electric mobility, as electric cars, as well as their charging infrastructure [78], will be networked, facing their own cybersecurity risks combined with a high amount of energy consumption under their control.

*3.2. Attack Scenarios*

Different vectors can be used to attack distribution and transmission systems to disrupt vital control systems. We assume an attacker accessing a PCN to aim at disrupting the power grid and do not specifically consider pure passive attacks, such as industrial espionage. In the following, we briefly discuss the three most important methods an attacker with access to the PCN can employ.

3.2.1. Disconnecting Resources

If an attacker has gained full access to the PCN, we can assume that the attacker is able to send arbitrary control commands to connected control systems. This allows an attacker, e.g., to control switches in substations which disconnect entire power lines or power plants from the grid, possibly leading to an immediate loss of the energy supply to consumers. In the attack on Ukraine in 2015, 225,000 consumers were disconnected from the grid, as attackers were able to control switches in multiple substations [21].

3.2.2. Injecting False Information

If an attacker can only gain control over a small subset of field devices, he can still indirectly influence the power grid, e.g., by sending forged or manipulated sensor readings to the control room [25,79,80]. The operators in the control room may act on the wrong data and take steps to correct a non-existent problem [81,82], which may lead to disruption in the power grid, e.g., unintentionally overloading a power line because sensor readings show a normal load. Depending on the sophistication of the attack, SCADA software may be able to identify a problem through bad data detection algorithms. However, these have been shown to not always be effective [83].

3.2.3. Denial of Service

Even if attackers neither have full access to the PCN nor can inject (false) information, they may still be able to manipulate certain devices and effectively render them non-functional to launch a denial of service attack against parts of the power grid [25]. For example, in the attacks on the Ukraine power grid in 2016 [22], the CRASHOVERRIDE malware used by the attackers was able to disable Siemens SIPROTEC devices, manipulate the firmware of serial-to-Ethernet devices, and hence disrupt crucial substation functions, such as protection, automation, or measurement. In future attacks, similar methods could, e.g., allow attackers to overload power lines even if these are secured by protection devices, potentially leading to physical damage.

In addition, from a communication networks perspective, power grids and their PCNs can be prone to denial of service attacks on their communication infrastructure, potentially disrupting vital control systems [84,85]. In such a denial of service attack on the communication infrastructure of power grids, an attacker strives to overload critical systems, e.g., a VPN entry point for remote maintenance (cf. Section 2.1), by creating an excessively high load [86], e.g., by performing a Ping of Death, TCP SYN flood, or UDP flood attack [87].

In a distributed denial of service attack, the malicious traffic originates from various sources, e.g., an Internet-wide botnet under the control of the attacker (cf. Section 3.1.7) relying on attack tools, such as Trinoo, Tribe Flood Network, or Stacheldraht [88]. To further amplify the impact of an attack with given resources, a distributed reflected denial of service attack sends requests with forged origin information to benign Internet services, such as DNS and NTP, with the goal to overload the target system with the corresponding responses [86]. Different works have shown that protocols typically used in PCNs of power grids, such as DNP3 or IEC 60870-5-104 (cf. Section 2.1), as well as different components found in such PCNs, are prone to various kinds of denial of service attacks [43,85,89].

## 4. Providing Cybersecurity for Interconnected Power Grids

Given the tremendous threats resulting from the diverse set of attack vectors and scenarios, providing security for power transmission and distribution within the grid as a critical infrastructure is a paramount objective. In our perspective, future improvements in the security of power grids will have to be a combination of technical approaches, awareness measures, and closer collaboration between the electrical engineering community and cybersecurity experts. To provide a way forward for security in interconnected power grids, thus, we identify a set of diverse security solutions and approaches, including both security software, as well as organizational measures, such as security training, which complement each other nicely. More specifically, we draw from the principle of *defense-in-depth* [90–92], as illustrated in Figure 2, to provide a comprehensive set of security measures at different layers. These measures encompass approaches for (i) device and application security, (ii) network security, (iii) physical security, and (iv) policies, procedures, and awareness. While discussing all four aspects in more detail in the following, we specifically focus on approaches to provide network security, since the need to protect communication becomes especially important with an increasing interconnection of power grids.

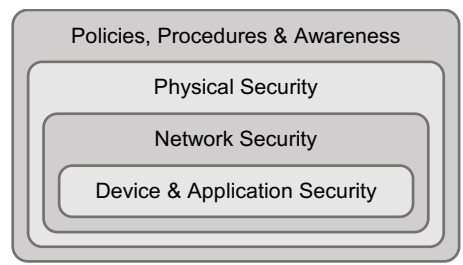

**Figure 2.** Following the principle of defense-in-depth, providing security for interconnected power grids needs to encompass a comprehensive set of measures for (i) device and application security, (ii) network security, (iii) physical security, as well as (iv) policies, procedures, and awareness.

*4.1. Device & Application Security*

As a foundation to provide defense-in-depth, all devices and applications deployed in power grids have to be secured. This becomes especially important considering an increasing interconnection in power grids, thus exposing potentially vulnerable devices and applications to larger attack surfaces. Especially, in the context of interconnected power grids, one promising approach consists of device and application diversity, i.e., using a wide range of different hardware and software to thwart a worst-case scenario where malware pervasively compromises all equipment [93]. Before integrating new devices into the network or deploying updates to devices, static firmware analysis [94] can be used to detect susceptibility to vulnerabilities, as well as malware and default credentials contained in a firmware image. Following a different direction, security assessments allow to additionally detect insecure configurations, as well as test for vulnerabilities at runtime [95–97]. Such approaches also show promising benefits for device and application security in power grids [98–100].

Once firmware has been analyzed, assessed for security, and subsequently deployed on devices controlling the power grid, such as PLCs, operators have to ensure that the appli-cations running on these devices do not get compromised. To this end, different approaches for hardware-based [101] or software-based [102] remote code attestation allow to verify the integrity of code, its execution, and updates. Within the context of interconnected power grids, remote attestation can, e.g., be used to detect compromised devices based on changes in their physical memory [93,103]. Likewise, hardware performance counters can be utilized to detect modifications in firmware of critical infrastructure components [104]. From a different angle, and complementing approaches to provide network security, deploying security agents to individual components, such as PLCs, aids in applying security patches, and maintaining end-to-end-security, as well as managing alarms [105–107].

*4.2. Network Security*

The increasing use of digital technology in power grids demands for more and more networking capabilities, resulting in the connection of previously isolated components to larger communication networks [12,14]. To address resulting security concerns, we require methods, to proactively prevent security incidents, most prominently using *network separation*, as well as to detect security incidents that successfully bypassed preventive measures using *intrusion detection systems*.

4.2.1. Network Separation

Traditional network separation through demilitarized zones (DMZ) and virtual networks is a standard tool for securing networks. These techniques make it harder for attackers to get a comprehensive view of the network through simple reconnaissance methods and restrict lateral movement within the network. DMZ and virtual networks are already used today by grid operators [98], and their deployment is often enforced through laws [108]. More recently, software-defined networking (SDN) has been providing a more flexible alternative to DMZ and virtual networks, which are usually configured once at the creation of a network and cannot easily be changed during operation. With SDN, changes to network separation can be configured quickly, e.g., to counter cyber attacks. Likewise, SDN can be used to enforce network compliant behavior, e.g., specified using communication rules by the operator of the communication network [109,110]. Dong et al. [111] and White et al. [112] provide more information on the opportunities and challenges of using SDN for smart grid resilience. The main challenge of SDN approaches for securing power grids result from availability concerns of grid operators.

4.2.2. Intrusion Detection Systems

Intrusion detection systems (IDS) are used in most company networks to detect attackers through suspicious network activities [113,114]. While not all companies in the energy sector are using IDS to secure their PCNs today, often citing availability concerns, we believe that IDS are especially well-suited to provide security in interconnected power grids: In contrast to office networks of large companies, traffic in PCNs is well defined as only certain protocols are used and, in most cases, any piece of software or hardware communicating over the network is known in advance [115]. An IDS in a PCN can, therefore, be much more restrictive without affecting operations than in an office network, where a large number of protocols and devices might communicate with each other and unknown Internet endpoints. Still, an attacker doing active reconnaissance (e.g., a port scan) could even easily be detected by a simple traditional IDS which is not specialized on industrial control systems.

Existing IDS deployments can be classified into network-based and host-based approaches [91]. In our opinion, both approaches should be used to secure PCNs. Furthermore, distributed and process-aware IDS allow to further strengthen security in interconnected power grids. In the following, we discuss the different approaches for IDS and their application to the energy sector.

Most of the communication in a PCN in the energy sector is conducted between the control room and substations or field devices. Consequently, a *network-based IDS (NIDS)* should be deployed at nodes in the network, such as switches, to be able to monitor any traffic within the PCN, especially those sent from the control room or towards it. Such an IDS can then easily detect reconnaissance measures, e.g., port scans. Moreover, as traffic in PCNs is well defined, a network-based IDS can detect further suspicious network activity, e.g., an increase in the number of packets sent from a network node or communication between network nodes that have not communicated beforehand. An emerging challenge in NIDS within a PCN is the question of how to cope with the ever changing and dynamic smart grid environment [79,113].

A *host-based IDS (HIDS)* runs on a single host in a networked environment and is, therefore, only able to monitor the incoming and outgoing traffic on this device. However, in contrast to NIDS, HIDS also monitor log files, system policies, and other relevant data concerning the host it is installed on [116]. These systems should be used wherever possible to supplement network-based IDS in the PCN. Not all switches in a network may offer a monitoring port or there might be other restrictions, why network-based systems cannot be used for parts of the network. In such cases, host-based IDS can be used to still monitor most parts of the traffic. Host-based IDS can, e.g., detect and block a denial-of-service attack targeting a device within the PCN. However, within a PCN a challenge for HIDS are the specialized hardware used. It will only be possible to install host-based IDS on a certain subset of systems. For highly at-risk hardware, e.g., some PLCs, a host-based IDS could be implemented by using a gateway between the network and the PLC to monitor any incoming and outgoing traffic [117].

A *distributed IDS* for the power grid combines network-based and host-based IDS and centrally aggregates and correlates data provided by the systems distributed within the PCN [118]. Such a distributed IDS allows a more complete view on the traffic within the PCN compared to many individual systems. Attacks from multiple points within the network can be correlated and dealt with more specifically. Moreover, compromised devices may be more easily identified. The collected data can be processed at a central location and could be used to help the operators in the control room to make informed decisions. A major challenge for distributed IDS within a PCN is the placement of the nodes as this influences the quality of the aggregated data.

A *process-aware IDS* employs context information about the environment in which it is placed. Chromik et al. [119] recently developed such a system for local substation networks. In their approach, which employs the event-based network security monitor Bro (now Zeek) [120], they check incoming control commands for consistency with safety requirements and physical constraints. Their approach maintains a model of the local system, which is updated through sensor readings and commands sent through the IEC 60870-5-104 protocol. At the current stage, inconsistent commands yield an alert message, which has to be acted on manually. Such systems could, e.g., be extended to block commands if their execution would lead to an inconsistent state, especially when adapted to more modern grid communication protocols.

The concept of process-aware IDS is especially interesting in a SCADA and industrial control systems (ICS) environment, where valid traffic is well-defined and commands lead to physical changes in the controlled system [121]. In combination with a distributed IDS, such approaches can lead to substantially more secure power grids, while introducing a minimum of extra hardware into the PCN. However, the main challenge in deploying such a system lies in the considerable knowledge which is required about the individual environment. Future work is required to automatically adapt to the differences between grid operators, as the used hardware, software, and protocols will be different.

In contrast to an IDS, which targets to detect attacks, an *intrusion prevention system (IPS)* operates similarly to identify attacks but, instead of merely raising an alarm, directly blocks any detected suspicious (network) activity and, thus, strives to prevent (and not only detect) potentially disastrous attacks [122]. While IPS appear promising to secure power grids as they can be used to automatically thwart attacks [123], they come with the risk of mistakenly blocking legitimate, potentially safety-critical communication as it is prevalent in PCNs. For example, an IPS might falsely classify seldom disturbance scenarios in a substation protection system [124] as attacks and, thus, prevent important reactive measurements from the control room. Consequently, automatically blocking potentially suspicious activity, e.g., using an IPS, should be restricted to non safety-critical communication. However, when applied to, e.g., a VPN entry point used for remote maintenance (cf. Section 2.1), IPS can nicely complement other security measures with their ability to directly block suspicious activity.

*4.3. Physical Security*

As grid operators often use their own physical networks for communication, physical security is directly related to cybersecurity. For example, a motivated attacker could break into a substation and infect local devices with malware or otherwise tamper with the available access to the PCN. Physical security for substations differs widely between grid operators but might be as low as a wire fence and no means of surveillance or access control [58]. Recommended measures to ensure physical security for substations include the protection of information on substations, such as engineering drawings and power flow models, and surveillance and monitoring measures, such as video cameras and motion detectors, as well as the restriction of physical access [59].

More sophisticated physical security may not only deter attackers but also act as a part of a general IDS. For example, if a physical security violation is detected at a substation [125], subsequent malicious activities in the PCN could be correlated with the physical security violation, helping in attack response. A major challenge will be integrating (automatically) detected physical security violations [125] into an overall security solution involving IDS on the network side. With an increasing integration of novel, easily-accessible assets, such as smart meters [16] and charging infrastructure for electronic vehicles [78,126], into the communication infrastructure of smart grids, the challenge of physical security is further exaggerated.

*4.4. Policies, Procedures & Awareness*

To date, the most devastating cyber attacks on power grids all specifically exploited human behavior either through spear-phishing, most prominently using emails, or manipulated downloads [21,22]. These problems cannot solely be solved by using more sophisticated security technology. Consequently, employees need to be trained to increase awareness towards security-related behavior [127].

Especially, workers who have direct access to vital equipment need to be aware of social engineering techniques and empowered to detect simple attacks, such as spear-phishing. For example, phishing experiments are valuable to raise employees' awareness for spear-phishing at companies in the electrical power domain [128]. However, as long as office networks and operational networks are not completely separated (which might be difficult to achieve), a general increase in security awareness is necessary to increase overall security. A challenge of security-related awareness training is its adaptation to the constantly changing cyber attack landscape.

Additionally, even with the best security measures and awareness trainings in place, eventually a (potential) security incident will occur [129]. As such, corresponding incident response plans and guidelines need to be created, maintained, and trained. However, most existing guidelines for responding to security incidents are mainly concerned with information chains and organizational processes [130,131] and not with actually remedying security incidents.

Consequently, grid operators need to develop and maintain actionable incident response plans and guidelines, supporting their employees with precise instructions also at the technical level on how to react to security incidents. To further decrease incident response times, it is recommended to keep the number of involved parties small, e.g., by operating own networks, andthus , thus, reduce the need for synchronization and communication during incident response [129]. Notably, proper preparation for cybersecurity incidents also requires to regularly train the response to cybersecurity incidents, e.g., using a corresponding training simulator [132,133]. To be valuable for training employees of grid operators, such training environments need to closely model typical process control network as found in power grids [17,134].

## 5. Conclusions

With increasing digitization and decentralization, grid operators are faced with rapid changes in the amount of communication necessary and how this communication is conducted. As a result, more networking is introduced, creating a wider attack space for attackers. In this paper, we highlighted resulting fundamental security problems and attack vectors, which still have to be addressed in the coming years in order to maintain a high level of security and availability of power grids as a critical infrastructure [2–5].

To provide security in interconnected power grids, we discussed a set of diverse security solutions and approaches. Depending on the country and the specific grid operation company, current security measures range from non-existent to state-of-the-art. However, even if attackers are only able to control a small fraction of the power connected to the grid, they can still leverage mechanisms inherent to today's large power grids to cause considerable damage. Thus, only an overall increase in the security of a country's power grid provides an effective defense against sophisticated attacks. Consequently, we identified a combination of software and organizational approaches, including intrusion detection systems, software-defined networking, and awareness training, as promising candidates for achieving this goal.

With the increasing move towards smart grids as, e.g., evidenced by the advancing deployment of smart meters [16] and charging infrastructure for electronic vehicles [78,126], further security challenges, but also research trends and cyber innovations addressing these challenges, arise [26]. Currently, one of the most prominent research fields in smart grid security is the challenge of appropriately employing blockchain and distributed ledger technology, and the research community addresses this challenge from various angles. With a focus directly on smart grid security, approaches promise to realize distributed security for smart grids at a national and international level [135], secure the supply chain for assets in the grid [136], automate the verification of assets in the grid [137], or persistently store security audits [138]. Furthermore, different streams of research address the challenge of decentralized management and control of the grid, e.g., by realizing flexible, decentralized local energy markets to securely and efficiently integrate renewable energies into today's rather static energy markets [139,140] or by forecasting demands for the grid and self-adjusting the consumption of power based on surges or drops in energy prices [137]. However, to date, it remains unclear whether and to what extent blockchain and distributed ledger technology indeed can capitalize on the promise to strengthen cybersecurity in power grids, especially considering strict regulatory requirements for critical infrastructures, such as power grids [137].

The cyber landscape within power grids as a critical infrastructure [2–5] is drastically changing: To continually provide ubiquitous power, new cybersecurity threats have to be taken into consideration and protected against. Achieving these goals requires tight collaboration between cybersecurity experts and grid operators to develop and implement cybersecurity solutions that are tailored to the unique requirements of power grids. To this end, our theoretical contributions consolidated in this perspective paper provide the foundation for deeper practical research and experimental studies to pave the way forward to provide a high level of cybersecurity for interconnected power grids.

**Author Contributions:** Conceptualization, T.K., R.E., B.K. and M.H.; methodology, T.K. and B.K.; investigation, T.K., B.K. and I.H.; writing—original draft preparation, T.K.; writing—review and editing, M.H., R.E., B.K. and I.H.; visualization, T.K.; supervision, R.E. and M.H. All authors have read and agreed to the published version of the manuscript.

**Funding:** This research received no external funding.

**Institutional Review Board Statement:** Not applicable.

**Informed Consent Statement:** Not applicable.

**Data Availability Statement:** Not applicable.

**Conflicts of Interest:** The authors declare no conflict of interest.

## Abbreviations

The following abbreviations are used in this manuscript:

| | |
|---|---|
| CIA | Confidentiality, integrity, and availability |
| DB | Database |
| DMZ | Demilitarized zone |
| DSO | Distribution system operator |
| HIDS | Host-based intrusion detection system |
| HMI | Human-machine interface |
| ICS | Industrial control systems |
| IDS | Intrusion detection system |
| IPS | Intrusion prevention system |
| NIDS | Network-based intrusion detection system |
| PCN | Process control network |
| PLC | Programmable logic controller |
| SCADA | Supervisory control and data acquisition |
| SDN | Software-defined networking |
| TSO | Transmission system operator |
| VPN | Virtual private network |